\documentclass[journal, a4paper]{IEEEtran}
\usepackage{cite}
\usepackage{xcolor}

\usepackage{verbatim}

\usepackage{bm}
\usepackage{array}
\usepackage{wrapfig}
\usepackage{tabularx}
\usepackage{amsmath,epsfig,graphicx}
\usepackage{subfigure}
\usepackage{color}
\usepackage{colortbl}
\usepackage{amssymb}
\usepackage{amsfonts}
\usepackage{multirow}
\usepackage{hhline}
\usepackage{makecell}
\makeatletter
\def\hlinewd#1{%
\noalign{\ifnum0=`}\fi\hrule \@height #1 \futurelet
\reserved@a\@xhline}
\usepackage{booktabs}
\usepackage{algorithmic}
\usepackage{algorithm}
\usepackage{array}
\DeclareMathOperator*{\minimize}{minimize}

\newcommand{\figref}[1]{Fig. \ref{#1}}
\newcommand{\tabref}[1]{Table \ref{#1}}
\newcommand{\equref}[1]{(\ref{#1})}
\newcommand{\secref}[1]{Sec. \ref{#1}}
\newcommand{\algref}[1]{Alg. \ref{#1}}

\renewcommand{\algorithmicrequire}{\textbf{Input : }}
\usepackage{url}

\usepackage{etoolbox}
\makeatletter
\patchcmd{\@maketitle}
{\addvspace{0.5\baselineskip}\egroup}
{\addvspace{-2\baselineskip}\egroup}
{}
{}

\title{Optimal Allocation of ESSs in Active Distribution Networks to achieve their Dispatchability}
\author{Ji Hyun~Yi,~\IEEEmembership{Student Member,~IEEE,}
        Rachid~Cherkaoui,~\IEEEmembership{Senior Member,~IEEE,}
        and~Mario~Paolone,~\IEEEmembership{Senior Member,~IEEE}% <-this % stops a space
}

\begin{document}
	\maketitle
\begin{abstract}
This paper presents a method for the optimal siting and sizing of energy storage systems (ESSs) in active distribution networks (ADNs) to achieve their dispatchability. The problem formulation accounts for the uncertainty inherent to the stochastic nature of distributed energy sources and loads. Thanks to the operation of ESSs, the main optimization objective is to minimize the dispatch error, which accounts for the mismatch between the realization and prediction of the power profile at the ADN connecting point to the upper layer grid, while respecting the grid voltages and ampacity constraints. The proposed formulation relies on the so-called Augmented Relaxed Optimal Power Flow (AR-OPF) method: it expresses a convex full AC optimal power flow, which is proven to provide a global optimal and exact solution in the case of radial power grids. The AR-OPF is coupled with the proposed dispatching control resulting in a two-level optimization problem. In the first block, the site and size of the ESSs are decided along with the level of dispatchability that the ADN can achieve. Then, in the second block, the adequacy of the ESS allocations and the feasibility of the grid operating points are verified over operating scenarios using the Benders decomposition technique. Consequently, the optimal size and site of the ESSs are adjusted. To validate the proposed method, simulations are conducted on a real Swiss ADN hosting a large amount of stochastic Photovoltaic (PV) generation.
\end{abstract}
% \fi
\IEEEpeerreviewmaketitle
\section{Introduction}
\IEEEPARstart{P}{ower} balancing is becoming an increasingly challenging task due to growing volatility in power systems introduced by the high penetration penetration of non-dispatchable and stochastic generation.
 The spinning reserve requirement is growing to tackle not only the peak demand but also the stochasticity of \emph{prosumption}, which is defined as the load consumption minus the generated power, causing significant cost increases for both transmission system operators (TSOs) and distribution system operators (DSOs). In the distribution grid, the prosumption forecast uncertainties are originated not only by the load consumption but also by the distributed RES. Thus, the deviation of the network infeed from the prescheduled power intake from the grid, called \emph{dispatch error}, should be compensated by the system operating reserve, exposing DSOs to high imbalance cost. 
There has been increasing interest in using ESSs as a flexible resource to compensate for the dispatch error, thereby enhancing the \emph{dispatchability} of the ADNs and mitigate the aforementioned financial risk. The \emph{dispatchability} identifies the capability of a resource, or a network, to control the realized active power flow through the resource, or the network, to follow a pre-defined power schedule with high fidelity. Thanks to the ADN dispatchability, the responsible DSO ensures its profit by mitigating imbalance penalties \cite{koeppel2008improving}. In this respect, the optimal control strategies of ESSs to improve the dispatchability of the stochastic resource \cite{Eperez, cost_of_imbalance}, or distribution feeder \cite{dispatchability} has been addressed. 

However, the efficient deployment of ESSs significantly depends on the investment planning. The work in \cite{5648364} proposed sizing of a wind farm ESS to achieve its dispatchability. The ESS dispatch strategy is coupled with the assessment of its capacity and expected lifetime based on the confidence level of the power output w.r.t the predefined schedule. To ensure the economic optimality of decision-making, it is worth investigating the economic value of ESSs investments by taking into account their operational benefits and installation costs. The work in \cite{mitigate_risk} proposed a method for the optimal allocation of ESSs through a cost-benefit analysis considering mitigating the distribution utilities' risk for compensating the gap between the actual prosumption and the purchased energy from the forward market. \cite{zheng2017multi} is an extension of \cite{mitigate_risk}, where different agents, such as DSOs, wind farms, solar power stations and demand aggregators in ADNs, determine the capacity of ESSs based on the game theory approach to reduce the transaction costs risk due to the resources' forecast errors.

It is also worth observing that the capacity and placement of ESSs should comply with the characteristics of the chosen control strategy and the subsequent operational conditions of the system. In this regard, the optimal power flow (OPF) that accurately models operation and control of the ADN, should be embedded in the ESS planning tool. However, OPF-based ESSs planning is inherently burdensome to solve due to its non-convexity. There exists meta-heuristic methods such as fuzzy PSO algorithm used in \cite{mitigate_risk} and \cite{zheng2017multi}. However, these solution approaches cannot guarantee the global optimum or even a feasible solution. To tackle the non-convexity of the OPF problem, convexification approaches, such as the semi-definite programming (SDP) \cite{giannitrapani2016optimal} or the second-order cone programming (SOCP) relaxation \cite{6736137, spinning_reserve_generation_cost, grover2018optimal, Voldev_SOCP}, have emerged as a solid and rigorous alternative. 
The SOCP relaxation proposed in \cite{jabr2006radial} has been implemented for the optimal ESS allocation in radial grids due to its superior computational efficiency referring to the SDP relaxation. In \cite{spinning_reserve_generation_cost}, an ESS planning strategy was developed relying on the SOCP-OPF model with the objective of providing ancillary services to the TSO, and to cope with the wind variability. Authors of \cite{grover2018optimal} further utilized the SOCP relaxed model for the ESSs allocation and operation to minimize the grid losses and imported power in the ADN.  
The work in \cite{Voldev_SOCP} tackled the ESS planning and operation problem by decomposing it into two stages: first stage determines the total ESS size to prevent grid constraints violations due to PV power imbalance, and the second stage allocates ESSs with optimal sizes by employing SOCP-OPF to minimize the energy cost.  
%In the first stage, PV power imbalance that causes grid constraints violations is calculated a-priori to determine the total required ESS size. In the second stage, by employing SOCP relaxed OPF, the ESSs are allocated in the optimal sites with optimal sizes with respect to minimization of the energy cost. 
Meanwhile, the Authors addressed that the objective function and constraints should satisfy some necessary conditions to guarantee the exactness of the SOCP-OPF solution, and suggested a formula to verify the exactness a-posteriori. The main drawbacks were explicitly underlined in \cite{limit} by the fact that the exactness of the solution cannot be guaranteed especially in the presence of reverse line power flows and for the cases where the upper bound of nodal voltage and the line ampacity constraints are binding.This brings significant limitations on the applicability of the method to ADNs hosting distributed generation units with large capacities. Moreover, the model neglects the transverse elements of the lines, which can bring an infeasibility of the solution especially when ADNs are composed of coaxial underground cables. 
The work in \cite{nick_exact} solves this problem and proposed the Augmented Relaxed OPF (AR-OPF) to convexify the AC-OPF for radial grids. Their contribution demonstrates that, in the AR-OPF problem comprising an objective function strictly increasing with the grid losses, the conditions for the exactness of the solution are mild and hold for realistic distribution networks. The AR-OPF was implemented in the subsequent works on the optimal ESS planning problem while embedding grid reconfiguration with the objective of minimizing grid losses, voltage deviation and the line congestion \cite{nick_reconfiguration}.  

In this paper, we propose an operation-driven planning strategy of ESS to achieve the dispatchability of the ADN based on the AR-OPF model. The objective of achieving disatchability requires a substantial and non-trivial modification on the AR-OPF as well as on the solution approach proposed in \cite{nick_reconfiguration} in order to reach the exactness of the relaxed OPF. Moreover, we formulated the sizing problem into two blocks by modifying the objective term, constraints and variables related to the dispatch error. Meanwhile, we apply the Bender’s decomposition to handle the multi-layered decisions with numerous scenarios \cite{geoffrion1972generalized}. The contributions of the paper are two as follows.
%The contribution of the paper is that the allocation of ESS is determined for ADNs considering (i) the objective of achieving dispatchability of the ADN in presence of the uncertainty of prosumption (ii) while employing an exact convex model of OPF under mild conditions, by (iii) proposing a reformulation of the the structure of planning problem and the objective function to maintain the exactness of the OPF relaxation.
\begin{enumerate}	
\item The optimal allocation of ESS is determined based on an exact convex model of the OPF to address the dispatchability of the ADN in the presence of prosumption uncertainty, while accurately reflecting the operational condition of the ADN.
\item The structure of the planning problem and the objective functions are formulated accounting for the necessary conditions to guarantee the optimality and the exactness of the OPF relaxation.
%reformulated to accomplish the minimization of the dispatch error while maintaining the exactness of the OPF relaxation.
\end{enumerate}
The paper is organised as follows: in Section II, we introduce the structure of the optimization problem and explain the key parts in detail. In Section III, the proposed problem formulation and solution approach are described. Section IV contains a detailed application example referring to the planning of ESSs into a real ADN. Finally, Section V concludes the paper.
\section{Problem Structure}
%In this section, % the key elements composing the proposed method are explained. To begin with, 
%we introduce the formulation for achieving the dispatchability of ADN, embedded in a daily OPF problem. Then, we describe the AR-OPF model that accurately models the power flow and grid operational constraints. 
\subsection{System description}\label{sec:system}
Buses excluding the slack bus in an ADN are indexed with $l \in \mathcal{L}$. Lines whose downstream bus is bus $l$ is are also indexed with $l \in \mathcal{L}$. In this paper, optimal allocation of ESS is determined based on the operation of the ADN over the planning horizon $Y$, assuming that there is no change in system conditions over years. Each year, we classify days into day-types indexed with $d \in \mathcal{D}$. The uncertainty in prosumption in each day-type is represented by operating scenarios indexed with $\phi\in\Phi_d$ with probability $\lambda_{\phi d}$ for each scenario. The dispatch interval is identified by the index $t\in\{1,...,T\}=\mathcal{T}$, where $T$ corresponds to the scheduling horizon of the problem. Time indices are separated by a constant timestep $\Delta t$. The dispatch problem at day $d$ accounts only for the active power. Each node of the ADN has a non-dispatchable aggregated prosumption ($s_{l\phi t}$) defined at each scenario and time interval. The aggregated ADN active powers through the grid connecting point (GCP) to an upper layer grid in all scenarios ($P_{1\phi t}, \forall \phi\in\Phi_d$) are expected to follow a day-ahead determined daily dispatch plan at each time interval ($DP_{td}$) derived with the support of a forecasting tool. Then, at each node where an ESS is allocated (i.e., $U_l=1$, where $U_l\in\{0, 1\}$ is installation status of the ESS at bus $l$), it is dispatched at each scenario and time interval according to active power ($p^E_{l\phi t}$) and reactive power ($q^E_{l \phi t}$). The dispatched active power compensates for the gap between active power infeed and the dispatch plan, which results from the deviation of realized prosumption from the prosumption prediction. Consequently, the observed dispatch error at GCP can be minimized.
% , which signifies the gap between the dispatch plan and the realized active power infeed at GCP. This is defined as \emph{dispatch error}.
 %Therefore, the capability of compensating the dispatch error is determined with the capacity of ESS in energy reservoir and power rating. 
In summary, the ESS allocation problem to achieve the dispatchability of the ADN under study is a two-stage decision process: the first stage which deals with the binary decision variables on the location of the ESS ($U_l$) and the continuous decision variables on the capacity of the ESSs energy reservoirs ($C_l$) and their power rating ($R_l$), and the second stage which deals with daily dispatch problems, determining the decision variables on the ESSs active and reactive power for all operating scenarios. 
\subsection{Dispatch plan of the distribution feeder}\label{sec:achieving dispatchability}
%In the day-ahead scheduling, a scheduling of the active power through GCP is determined by the distribution grid operator with respect to the prosumption prediction. Meanwhile, ESSs are supposed to be exploited to cover the uncertainty of the prosumption. 
The operational benefit of ESSs allocation for the ADN dispatchability is evaluated with sets of operationg scenarios, where each set refers to a typical day. The set of day-types is supposed to be pre-selected by the modeler.  

The advantage of using scenarios is related to the use of generic parametric and non-parametric distributions of prosumptions. In this paper, the active and reactive prosumption scenarios for each day are generated with the assumption that the prosumption profile follow a normal distribution. Therefore, the average of the prosumption over the scenarios is equal to the given prosumption prediction, and the losses predictions are calculated by averaging the losses over the scenarios \cite{dispatchability}. It is understood that the variables with subscript $l, \phi, t$ are defined for $l\in\mathcal{L}, \phi\in\Phi_d, t\in\mathcal{T}$, respectively. $p_{l\phi t}$ is aggregated active prosumption at bus $l$, scenario $\phi$, time $t$, and $f_{l\phi t}$ is square of current magnitude causing losses in line $l$, scenario $\phi$, and time $t$. $r_l$ is the total longitudinal resistance of line $l$. $\tilde{p}_{ltd}$ and $\tilde{f}_{ltd}$ are avereage of the active load and squared current causing losses over all scenarios at bus $l$, time $t$, and day $d$, respectively. $\Delta p_{l\phi t}$ and $\Delta f_{l\phi t}$ are deviation (we would call it as \emph{error} hereafter) of prosumption and squared current causing losses for scenario $\phi$ and time $t$ from $\tilde{p}_{ltd}$ and $\tilde{f}_{ltd}$, respectively. \equref{eq:prosumption prediction} and \equref{eq:losses prediction} express the prosumption at each bus $l$ and the line losses of the line $l$ by these variables.   
In \equref{eq:dispatch plan}, a daily dispatch plan $DP_{td}$ follows the predicted point of the total prosumption considering the losses.
\begin{align}
& p_{l \phi t}=\tilde{p}_{ltd}{-}\Delta p_{l \phi t}, \forall l,\forall \phi, \forall t \label{eq:prosumption prediction}\\
& r_{l}f_{l \phi t}=r_{l}\tilde{f}_{ltd}{-}r_{l}\Delta f_{l \phi t}, \forall l,\forall \phi, \forall t  \label{eq:losses prediction}\\
& DP_{td}={\sum_{l\in \mathcal{L}}(\tilde{p}_{ltd}{+}r_{l}\tilde{f}_{ltd})}, \forall t \label{eq:dispatch plan}
\end{align}
In this context, the \emph{dispatch error} with no dispatchable resources in ADN is formally defined as the total error of prosumption plus the line losses over the buses/lines as shown in the left-hand side of \equref{eq:error expression}. The installed ESSs in ADN can be dispatched with acive power ($p_{l \phi t}^{E}$) to compensate for the error. 
%Therefore,  is equivalent to the compensated error at each bus $l$ by the ESSs located at each bus $l$, whereas. 
 $\epsilon_{l \phi t}$ represents the residual dispatch error that cannot be covered at bus $l$, scenario $\phi$, and time $t$. To quantify the covered or not covered error by the ESSs, we say that the sum of the two parts, shown in the right-hand side of \equref{eq:error expression}, is equal to the dispatch error in case of no dispatchable resources.  
\begin{align}\label{eq:error expression}
\sum_{l\in \mathcal{L}}(\Delta {p}_{l \phi t}{+}r_{l}\Delta f_{l \phi t})=\sum_{l\in \mathcal{L}}({\epsilon}_{l \phi t}{+}p_{l \phi t}^{E}), \forall \phi, \forall t
\end{align}
Finally, the objective term to minimize the power deviation from the daily dispatch plan for all scenarios and time intervals is expressed as follows.
\begin{align}\label{eq:modified dispatchability}
\minimize_{p^{E}} \sum_{t\in \mathcal{T}}\sum_{\phi \in \Phi_d}\lambda_{\phi d}|\sum_{l\in \mathcal{L}} \epsilon_{l \phi t}|
\end{align}
Once the OPF problem including the constraints related to the dispatchability is solved, we can calculate the capability of ADN with the allocated ESS sizes to reduce the dispatch error. In this context, we introduce a dispatchability index called \emph{leftover dispatch error rate} (LDER), and defined as $\theta_{\phi td}$. It represents the ratio between the resulting dispatch error and the anticipated dispatch error in case of no ESS at scenario $\phi$ and time $t$ for the daily operation on day $d$, as expressed in \equref{eq:error reduction rate_def} (* here indicates that it is the identified optimal solution). {In the denominator, indicating the dispatch error without ESS, the error regarding the grid losses is ignored because its magnitude is negligible compared to that of the prosumption error. 
\begin{align}
\label{eq:error reduction rate_def}
    \theta_{\phi td}=\frac{|\sum_{l\in\mathcal{L}}\epsilon_{l \phi t}^{*}|}{|\sum_{l\in\mathcal{L}}\Delta p_{l \phi t}|},\forall \phi, \forall t
\end{align}  
\subsection{Augmented Relaxed Optimal power flow}
\label{sec:AR_OPF}
For a radial power network, the power flow equations are given in \equref{eq:complex power balance up}-\equref{eq:relaxed internal current}. The variables with subscript $l$ are defined for $l \in \mathcal{L}$. The upstream bus of bus $l$ is notated as $up(l)$. $\mathbf{G}$ is the adjacency matrix of the network, where $\mathbf{G}_{k,l}$ is defined for $k,l\in\mathcal{L}$ and $\mathbf{G}_{k,l}=1$ if $k=up(l)$ or 0 if not. $S_l^t=P_l^t+jQ_l^t$ indicates the complex power injected into line $l$ from node $up(l)$. $S_l^b=P_l^b+jQ_b^t$ is the complex power injected into node $l$ from line $l$. $s_l=p_l+jq_l$ represents the given prosumption at node $l$. $v_{up(l)}$ and $v_l$ represents the nodal voltages at node $up(l)$ and $l$, respectively, whereas $f_l$ is the square of the longitudinal (i.e., internal) current through line $l$ which produces grid losses through the line longitudinal impedance $z_l=r_l+jx_l$. $b_l$ indicates the shunt susceptance of line $l$. $S_l^t$ and $S_l^b$ are determined by \equref{eq:complex power balance up}, and \equref{eq:complex power balance down}, respectively. \equref{eq:nodal voltage} determines the magnitude of nodal voltages, where $\mathfrak{R}(.)$ represents the real parts of a complex number. The current that produces losses at line $l$ is originally defined to be equal to the right-hand side of \equref{eq:relaxed internal current}. However, the equality is replaced with an inequality, as shown in \equref{eq:relaxed internal current} by applying the SOCP relaxation \cite{jabr2006radial}. It is noteworthy that the equality of \equref{eq:relaxed internal current} only holds when the terms in the objective function are strictly increasing with respect to the internal current $f_l$.
\begin{align}
& S_l^t=s_l{+}{\sum_{m\in\mathcal{L}}\mathbf{G}_{l,m}S_l^t}{+}z_l\,f_l{-}j(v_{up(l)}{+}v_l)b_l, \forall \l \label{eq:complex power balance up} \\
& S_l^b=s_l{+}\sum_{m\in\mathcal{L}}\mathbf{G}_{l,m}S_l^t, \forall \l \label{eq:complex power balance down} \\
%\end{equation}
%\begin{equation}
& v_l=v_{up(l)}{-}2\mathfrak{R}\bigg(z_l^*\Big(S_l^t{+}j v_{up(l)}b_l\,\Big)\bigg){+}|z_l|^2{f_l}, \forall \l \label{eq:nodal voltage} \\
%\end{equation}
%\begin{align}
& f_{l}v_{up(l)}\,\geq\,|S_l^t{+}j v_{up(l)}b_l|^2, \forall \l \label{eq:relaxed internal current}
\end{align}

In order to avoid any inexact solution (i.e., any solution that makes the left-hand side of \equref{eq:relaxed internal current} strictly greater than the right-hand side, and thus without physical meaning), the Authors of \cite{nick_exact} introduced auxiliary variables to formulate the AR-OPF model. These are $\Bar{f}_l$, $\Bar{S}_l=\Bar{P}_l{+}j\Bar{Q}_l$, $\Bar{v}_l$ and $\hat{S}_l=\hat{P}_l{+}j\hat{Q}_l$, which stand for the upper bounds for internal current, complex power, voltage and lower bounds for the complex power, respectively. $v^{max}$ and $v^{min}$ are upper bounds and lower bounds of the squared nodal voltage magnitude, respectively. $I^{max}_l$ indicates upper limit on the squared current of line $l$. $P^{max}_l$ and $Q^{max}_l$ are upper limits of active and reacitve power flows for line $l$, respectively.
 The branch power flow, nodal voltage, and current equations are defined as well with the set of the auxiliary variables, as shown in \equref{eq:lowerbound complex power balance up}-\equref{eq:upperbound internal current2}. \equref{eq:lowerbound complex power balance up} and \equref{eq:lowerbound complex power balance down} indicate the lower bound of branch power flow at the sending end and the receiving end of line $l$. The upper bound nodal voltage is determined correspondingly with \equref{eq:nodal upperbound voltage}. Likewise, the branch power flow equations for upper bound power flow variables are shown as \equref{eq:upperbound complex power balance up} and \equref{eq:upperbound complex power balance down}. \equref{eq:upperbound internal current} and \equref{eq:upperbound internal current2} express that the upper bound of the current $f_l$ should be decided by the maximum of absolute complex power flow from both sides of line $l$. The voltage constraint, the ampacity constraint from the sending end and the receiving end are modeled as in \equref{eq:voltage constraint}-\equref{eq:ampacity from downside}. Finally, \equref{eq:upperbound power range} are added to complete the set of equations required to guarantee the exactness. 
\begin{align}
& \hat{S}_l^t=s_l{+}\sum_{m\in\mathcal{L}}\mathbf{G}_{l,m}\hat{S}_l^t{-}j(\Bar{v}_{up(l)}{+}\Bar{v}_l)b_l, \forall \l \label{eq:lowerbound complex power balance up} \\
%\end{align}
%\begin{equation}\label{eq:lowerbound complex power balance up}
%\hat{P}_l^t=\tilde{p}'_l-\Delta p'_l{+}p^E_l{+}\sum_{m\in\mathcal{L}}\mathbf{G}_{l,m}\hat{P}_l^t, \ \forall \l
%\end{equation}
%\begin{equation}\label{eq:lowerbound complex power balance up}
%\hat{Q}_l^t=q_l{+}q^E_l{+}\sum_{m\in\mathcal{L}}\mathbf{G}_{l,m}\hat{Q}_l^t{-}(\Bar{v}_{up(l)}{+}\Bar{v}_l)b_l, \ \forall \l
%\end{equation}
%\begin{equation}
& \hat{S}_l^b=s_l{+}\sum_{m\in\mathcal{L}}\mathbf{G}_{l,m}\hat{S}_l^t, \forall \l \label{eq:lowerbound complex power balance down}\\
%\end{equation}
%\begin{equation}
& \Bar{v}_l=\Bar{v}_{up(l)}{-}2\mathfrak{R}\Big(z_l^*(\hat{S}_l^t{+}j\Bar{v}_{up(l)}b_l)\Big), \forall \l \label{eq:nodal upperbound voltage}\\
%\end{equation}
%\begin{equation}
& \Bar{S}_l^t=s_l{+}\sum_{m\in\mathcal{L}}\mathbf{G}_{l,m}\Bar{S}_l^t{+}z_{l}f_l{-}j(v_{up(l)}{+}v_l)b_l, \forall \l \label{eq:upperbound complex power balance up}\\
%\end{equation}
%\begin{equation}\label{eq:upperbound complex power balance up}
%\Bar{P}_l^t=\tilde{p}'_l-\Delta p'_l{+}p^E_l{+}\sum_{m\in\mathcal{L}}\mathbf{G}_{l,m}\Bar{P}_l^t{+}z_{l}f_l, \ \forall \l 
%\end{equation}
%\begin{equation}\label{eq:upperbound complex power balance up}
%\Bar{Q}_l^t=q_l{+}q^E_l{+}\sum_{m\in\mathcal{L}}\mathbf{G}_{l,m}\Bar{Q}_l^t{-}(v_{up(l)}{+}v_l)b_l, \ \forall \l 
%\end{equation}
%\begin{equation}
& \Bar{S}_l^b=s_l{+}\sum_{m\in\mathcal{L}}\mathbf{G}_{l,m}\Bar{S}_l^t, \forall \l \label{eq:upperbound complex power balance down}\\
%\end{equation}
%\begin{equation}}
& \begin{aligned}
\Bar{f}_{l}v_l & \geq|\max{\big\{|\hat{P}_l^b|,|\Bar{P}_l^b|\big\}}|^2\\
&{+}|\max{\big\{|\hat{Q}_l^b{-}j \Bar{v}_{l} b_{l}|,|\Bar{Q}_l^b{-}j v_{l}b_{l}|\big\}}|^2, \forall \l
\end{aligned} \label{eq:upperbound internal current} \\
%\end{equation}
%\begin{equation}
& \begin{aligned}
 \Bar{f}_{l}v_{up(l)}&\geq|\max{\big\{|\hat{P}_l^t|,|\Bar{P}_l^t|\big\}}|^2\\
&{+}|\max{\big\{|\hat{Q}_l^t{+}j\Bar{v}_{up(l)} b_{l}|,|\Bar{Q}_l^t{+}j v_{up(l)}b_{l}|\big\}}|^2, \forall \l 
\end{aligned} \label{eq:upperbound internal current2}\\
%\end{align}
%\end{equation}
%\begin{equation}
%\begin{align}
& v^{min}\,\leq\,v_l,~~ \Bar{v}_{l}\,\leq\,v^{max}, \forall \l \label{eq:voltage constraint} \\
%\end{equation}
%\begin{equation}
& I_{l}^{max}v_{up(l)}\geq|\max{\big\{|\hat{P}_l^t|,|\Bar{P}_l^t|\big\}}|^2{+}|\max{\big\{|\hat{Q}_l^t|,|\Bar{Q}_l^t|\big\}}|^2, \forall \l \label{eq:ampacity from upperside} \\
%\end{equation}
%\begin{comment}
%\begin{equation}\label{eq:ampacity from downside}
%I_{l}^{max}v_l\geq
%	\begin{aligned}
%			&|\max{\big\{|\hat{P}_l^b|,|\Bar{P}_l^b|\big\}}|^2\\
%		{+}	&|\max{\big\{|\hat{Q}_l^b|,|\Bar{Q}_l^b|\big\}}|^2
%	\end{aligned}, \ \forall \l
%\end{equation}
%\end{comment}
%\begin{equation}
& I_{l}^{max}v_l\geq|\max{\big\{|\hat{P}_l^b|,|\Bar{P}_l^b|\big\}}|^2{+}|\max{\big\{|\hat{Q}_l^b|,|\Bar{Q}_l^b|\big\}}|^2, \forall \l \label{eq:ampacity from downside} \\
%\end{equation}
%\begin{align}
& \Bar{P}_l^t \leq\ P_l^{max}, \qquad \Bar{Q}_l^t \leq\ Q_l^{max}, \forall \l \label{eq:upperbound power range}
\end{align}
For the sake of readability, the equations mentioned above are grouped and represented by $\Theta(\varphi,\kappa)\geq0$ where $\varphi=\{S^t,v,f,\hat{S}^t,\Bar{v},\Bar{f}, \Bar{S}^t,s\}$ is the set of variables and $\kappa=\{\mathbf{G},z,b,v^{max},v^{min},I^{max}, P^{max}, Q^{max}\}$ is the set of parameters. The notation without subscript corresponds to the vector of variables and parameters for all buses/lines. 

As already quantified in \cite{nick_exact}, the set of grid constraints employing the auxiliary variables slightly shrinks the original feasible solution space, removing solution space related to undesirable or extreme operation points of the network near the upper bound of nodal voltage or lines' ampacity limits.\footnote{We show that the compression of the solution space caused by the augmented constraints is small by following the steps of numerical analysis reported in \cite{nick_exact}. Under the case study shown in \secref{sec:simulation}, We make one of two operating constraints (voltage upper bound constraint and ampacity constraint) binding at one node or line and relax the other one to find the difference between the physical state variables (nodal voltage-magnitudes and original current flow) and corresponding auxilary variables. The difference between the nodal voltage-magnitude and the auxillary one is 0.001\%. The difference between the original current flow and the auxillary one is equal to 0.2\%.}
%The AR-OPF problem is defined as \equref{eq:objective_AR_OPF}, where $C(r_l f_l)$ is a cost function of the grid losses. 
% along with the mild exactness conditions defined with the grid parameters, which are elaborated in \cite{nick_exact} in detail.
%\begin{equation}\label{eq:objective_AR_OPF}
%\minimize_{s,S,v,f,\hat{S},\hat{v},\Bar{S},\Bar{f}} \sum_{l\in \mathcal{L}} C(r_l f_l)
%\end{equation}
%\begin{center}
%subject to: \equref{eq:AROPF_constraints}
%\end{center}
%\begin{center}
%\equref{eq:complex power balance up}-\equref{eq:nodal voltage},  \equref{eq:complex power balance down}, \equref{eq:relaxed internal current}, \equref{eq:lowerbound complex power balance up}-\equref{eq:upperbound reactive power range}
%\end{center}
\subsection{Energy storage systems}
\label{sec:ESS}
The ESS investment is modeled through \equref{eq:available p rating}-\equref{eq:c-rate}.  
 $\mathcal{C}^{max}_l$/$\mathcal{C}^{min}_l$ is maximum/minimum possible ESS energy reservoir capacity at bus $l$. $\mathcal{R}^{max}_l$ and $\mathcal{R}^{min}_l$ represent maximum and minimum possible ESS power rating capacity at bus $l$. In reality, available power ratings and energy capacities are often restrained as seen in \equref{eq:available p rating} and \equref{eq:available e capacity} due to various physical constraints involving, for instance, manufactural or geographical factors. $CR_{min}$ is the minimum value for the rate at which ESS is discharged relatively to its maximum energy capacity. The power rating and energy reservoir is determined considering this relationship as shown in \equref{eq:c-rate}. 
\begin{align}
%&\sum_{l \in \mathcal{L}}(\mathcal{I}_{c}U_l{+}\mathcal{I}_{p}R_{l}{+}\mathcal{I}_{e}C_{l})\leq \mathcal{B}, \ \forall \l \label{eq:budget limit}\\
%\end{align}
%\begin{equation}
& \mathcal{R}^{min}_{l}U_{l}\leq R_l\leq {R}^{max}_{l}U_{l}, \forall \l \label{eq:available p rating} \\ 
%\end{equation}
%\begin{equation}
& \mathcal{C}_l^{min}U_l\leq C_l\leq \mathcal{C}_l^{max}U_l, \forall \l \label{eq:available e capacity} \\
%\end{equation}
%\begin{align}
& R_l \leq \frac{C_l}{CR_{min}}, \forall \l \label{eq:c-rate}
\end{align}
The operational characteristic of ideal ESS is modeled with equations \equref{eq:power ESS}-\equref{eq:SOE limit}. The variables with subscript $l$ are defined for $l \in \mathcal{L}$. $E_{lt}^{E}$ is state-of-energy (SoE) of ESS installed at bus $l$ for time $t$. $E^{max}_l$/$E^{min}_l$ is maximum/minimum allowed SoE level. The capability curve of a given ESS is linearized by constructing an inscribed square within the original circular capability curve defined by the maximum complex power of the ESS (see \equref{eq:power ESS}). The SoE level of ESS changes with the charge/discharge of the ESS at every time interval, as described in \equref{eq:dynamic SOE}. Also, \equref{eq:SOE limit} indicates that the SoE of the ESS should be within the SoE limits. 
\begin{align}
& {-}\frac{R_l}{\sqrt{2}}\leq p_{lt}^{E}\leq \frac{R_l}{\sqrt{2}}, \quad {-}\frac{R_l}{\sqrt{2}}\leq q_{lt}^{E}\leq \frac{R_l}{\sqrt{2}}, \forall l, \forall t
\label{eq:power ESS}  \\  
%\end{align} \\
%\begin{equation}
& E_{l(t{+}1)}^{E}=E_{lt}^{E}{+}\Delta t*p_{lt}^{E}, \forall l, \forall t
\label{eq:dynamic SOE} \\
%\end{equation}
%\begin{equation}
& E^{min}_l C_l \leq E_{lt}^{E} \leq E^{max}_l C_l, \forall l, \forall t
\label{eq:SOE limit} 
%\end{equation}
\end{align}
In the interest of brevity, \eqref{eq:power ESS}-\eqref{eq:SOE limit} are indicated by $\Xi(\eta_t,\xi)\geq0, \forall t \in \mathcal{T}$ where $\eta=\{p^{E},q^{E},E^{E}, U, R, C\}$ is the set of variables and $\xi=\{\Delta t, E^{min}, E^{max}\}$ is the set of parameters. The notation without subscript corresponds to the vectors of variables and parameters for all buses/lines.
\section{Problem Formulation}
The objective of the problem is to determine the optimal sizes and sites of ESSs so that the active power through the GCP follows the dispatch plan with minimal deviation. However, it is clear that the dispatch error described as \equref{eq:modified dispatchability} in \secref{sec:achieving dispatchability} does not increase while the total grid losses increase (a necessary condition of the AR-OPF). Therefore, the exactness of the solution cannot be guaranteed if the objective value that corresponds to the objective term \equref{eq:modified dispatchability} is significant in magnitude compared to objective term regarding the total grid losses in the objective function of the AR-OPF model.
%The problem is formulated as a two-stage stochastic mixed-integer SOCP model. In the first stage, the site and size of ESSs are determined based on the investment cost. In the second stage, the expected penalty cost on the grid losses and the deviated energy from the dispatch plan during an operation of the system equipped with the allocated ESS is evaluated based on the generated prosumption scenarios.

Therefore, we propose to decompose the problem into two blocks each consisting of an OPF problem. In this way, we can exclude \equref{eq:modified dispatchability} from the AR-OPF problem and convey it to another, approximated, OPF problem (the so-called 1st block problem), which aims to find the optimal level of dispatchability based on the ESSs investment cost and the imbalance penalty.
% the first block where an approximated OPF formulation is considered.
 The determined dispatchability level, defined as LDER, is then introduced to an AR-OPF model-based problem (the so-called 2nd block problem) as an index with which the ADN has to comply with.
 % The objective of the 2nd block problem becomes a minimization of the total unserved load upto near zero or small enough value to ensure the feasible operational condition of the ADN, while the dispatchability constraint is satisfied.}  

The whole algorithm of the proposed approach is illustrated in \figref{fig:full_algorithm}. In the 1st block problem, the optimal ESSs allocation, the daily dispatch plans, and the corresponding LDER are calculated employing linear approximated OPF ignoring the grid losses. Only the nodal voltage constraints are considered regarding the operational constraints, ignoring the ampacity limits to reduce the computational burden. Afterward, the outputs of the 1st block, which are the ESS allocation and the LDER, are used as inputs for the 2nd block problem.  

In the 2nd block problem, the objective is to refine the optimal allocation of the ESSs, considering several operating scenarios and achieving the same level of LDER calculated in the 1st block problem. In this respect, LDER is implemented as an additional constraint to an AR-OPF model, which considers the full AC-OPF as well as voltage constraints and line ampacity limits. Then, the size and site of the ESSs are iteratively adjusted thanks to the Benders decomposition technique. This iterative process starts initially with a feasibility check of the ESS allocation resulting from the 1st block problem. 
\begin{figure}[h]
    \centering
    \includegraphics[width=6cm]{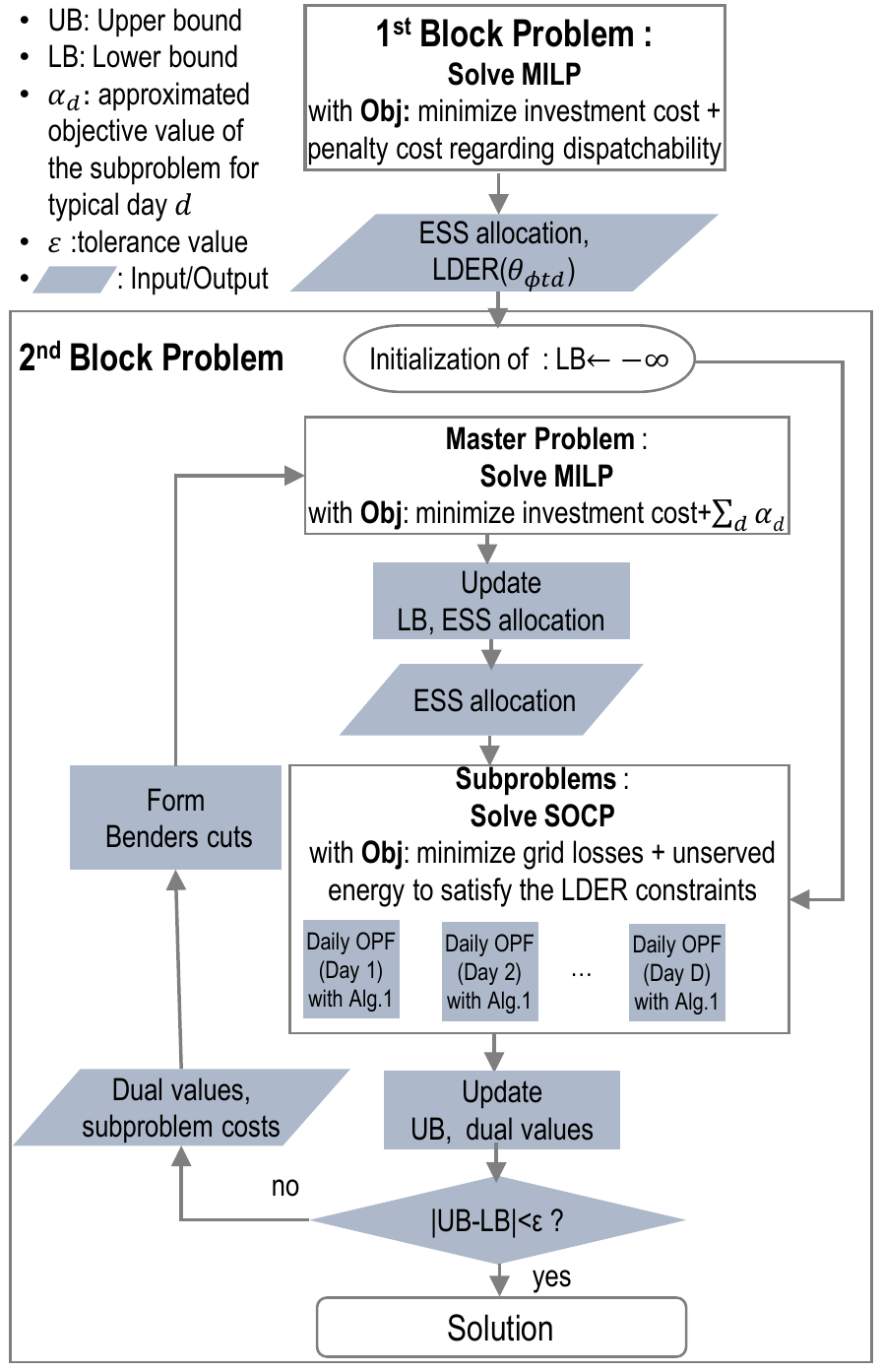}
    \vspace{-5pt}
    \caption{Full algorithm of the proposed method.}
    \label{fig:full_algorithm}
\end{figure}
\subsection{1st block problem}
We minimize the investment and total penalty costs over the planning horizon to find out the optimal allocation of the ESSs and the optimal dispatchability level. We embed an approximated OPF constraints for all operating scenarios into a two-stage mixed-integer linear programming problem (see \secref{sec:system}). The OPF is formulated by the linear Distflow model in which shunt elements are considered, whereas the grid losses are neglected. In this way, the reactive power generated by the shunt impedance of the lines is accounted for in the nodal voltage constraints. Meanwhile, neglecting the formulation of the squared current ($f_l$) (i.e., losses) is less likely to affect the feasible solutions in this stage since the ampacity constraint is ignored.
%Since the linearized Distflow is rigorously approximated model  we can embed the objective term to obtain optimal dispatch plan within the daily OPF. 
The dispatch plan follows the prosumption prediction, as shown in \equref{eq:distflow dispatch plan}, while the prosumption deviation at node $l$ is expressed as shown in \equref{eq:dist flow error expression}. We substitute \equref{eq:prosumption prediction} into the active power balance equation, resulting in \equref{eq:distflow power balance up dispatch}. The lossless Distflow power flow at both sides of line $l$ including the ESS power are expressed via \equref{eq:distflow power balance up dispatch}-\equref{eq:distflow reactive power balance down}. \equref{eq:distflow nodal voltage} calculates the nodal voltage, which is governed by voltage constraint as shown in \equref{eq:distflow voltage constraint}. All the variables within \equref{eq:dist flow error expression}-\equref{eq:distflow voltage constraint} are defined for $\l\in\mathcal{L}, \phi\in\Phi_d, t\in\mathcal{T}$, and $d\in\mathcal{D}$.
%; however the corresponding indices are omitted for the sake of brevity.
\begin{align} 
& DP_{td}=\sum_{l\in \mathcal{L}}\tilde{p}_{ltd}, \forall t, \forall d
\label{eq:distflow dispatch plan} \\
& \sum_{l\in L}\Delta p_{l}=\sum_{l\in \mathcal{L}}(\epsilon_{l}{+}p_l^E), \forall \phi, \forall t, \forall d 
\label{eq:dist flow error expression} \\
& P_l^t{=}P_l^b{=}\tilde{p}_l{-}\Delta p_l{+}p_{l}^{E}{+}\sum_{m\in\mathcal{L}}\mathbf{G}_{l,m}\,P_l^t,\forall l,\forall \phi, \forall t, \forall d
\label{eq:distflow power balance up dispatch} \\
& Q_l^t=q_l{+}q_{l}^{E}{+}\sum_{m\in\mathcal{L}}\mathbf{G}_{l,m}\,Q_l^t{-}(v_{up(l)}{+}v_l)b_l,\forall l,\forall \phi, \forall t, \forall d
\label{eq:distflow reactive power balance up} \\
& Q_l^b=q_l{+}q_{l}^{E}{+}\sum_{m\in\mathcal{L}}\mathbf{G}_{l,m}Q_l^t,\forall l,\forall \phi, \forall t, \forall d 
\label{eq:distflow reactive power balance down} \\
& v_l=v_{up(l)}{-}2\mathfrak{R}\Big(z_l^*(S_l^t{+}j v_{up(l)}b_l)\Big),\forall l,\forall \phi, \forall t, \forall d
\label{eq:distflow nodal voltage} \\
& v^{min}\,\leq\,v_l\,\leq\,v^{max},\forall l,\forall \phi, \forall t, \forall d
\label{eq:distflow voltage constraint}
\end{align}
 The objective function is defined as to minimize the investment cost (the first line of \equref{eq:objective function})  of ESSs and the annual penalty cost (the second line of \equref{eq:objective function}) regarding the uncovered dispatch error over the planning horizon $Y$. The penalty cost of day $d$ is the uncovered dispatch error over the operating scenarios for day $d$ multiplied by $\omega_d$, which is the cost coefficient for the imbalance. $N_d$ stands for the number of days in a year that have the same prosumption profile as that of typical day $d$. $\Omega_1$ and $\Omega_2$ represents the set of control variables in the first and second stage decision process, respectively. The constraints regarding the ESS allocation and operation explained in \secref{sec:ESS} are included (i.e., \equref{eq:available p rating}-\equref{eq:c-rate}, \equref{eq:ESS_constraints_1st}), along with the linear approximated lossless OPF constraints (\equref{eq:distflow dispatch plan}-\equref{eq:distflow voltage constraint}).
By solving the problem, the optimal allocation of ESSs along with the dispatch plan is obtained, and the daily LDERs are calculated by \equref{eq:error reduction rate_def} concerning all days with the index $d \in \mathcal{D}$.
\begin{align}
& \begin{aligned}
\begin{split}
\minimize_{\substack{\forall U,C,R \in \Omega_1;\\ \forall S^{t},v,s^{E}\in \Omega_2}}&\sum_{l\in \mathcal{L}} (\mathcal{I}_c U_l{+}\mathcal{I}_p R_{l}{+}\mathcal{I}_{e} C_{l})\\&{+}Y\sum_{d\in \mathcal{D}}N_{d}w_d\sum_{t\in \mathcal{T}}\sum_{\phi\in \Phi_d}\lambda_{\phi d}|\sum_{l\in \mathcal{L}}\epsilon_{l \phi t}|
\end{split}
\label{eq:objective function} 
\end{aligned}
\end{align} 
{\centering \text{subject to: \equref{eq:available p rating}-\equref{eq:c-rate}, \equref{eq:distflow dispatch plan}-\equref{eq:distflow voltage constraint}, } 
\begin{align}
\Xi(\eta_{\phi t d},\xi)\geq0, \ \forall \phi, \forall t, \forall d
\label{eq:ESS_constraints_1st}
\end{align}}
\vspace{-25pt}
\subsection{2nd block problem}
The dispatchability level is incorporated into the 2nd block problem as a constraint governed by the dispatchability index LDER. The objective of the 2nd block problem is to adjust the ESS allocation from the 1st block problem to the optimal site and size that can minimize the grid losses and unserved load. The system condition during the operation horizon is evaluated through solving the AR-OPF problem. Therefore, the 2nd block problem is formulated as a Mixed-integer second-order cone programming (MISOCP) problem. A convex SOCP model can be obtained from the non-convex MISOCP problem by using solution approaches that tackle the integer variables, such as the Branch and bound algorithm. However, the solution approach becomes computationally burdensome with the increase of integer variables. In this respect, we apply the Benders decomposition technique to decompose the 2nd block problem into a master problem and several parallel subproblems that each represents a daily OPF problem. The master problem determines the ESSs allocation, followed by the fitness evaluation of the determined allocations in the subproblems in terms of grid losses and unserved load. The unserved load takes values to ensure the feasibility of the subproblem regardless of the ESS allocation.

The initial step of the 2nd block is to solve parallel subproblems, checking the operational condition of ADN under the ESS allocation and the corresponding LDER calculated from the 1st block. The objective of the subproblem and the dual values of ESS allocation are computed and sent to the master problem to construct the first Bender's cut. Through multiple iterations between the master problems and the subproblems, the convergence is reached when the gap between the lower bound of the total cost (LB) and the upper bound of the total cost (UB) becomes less than a tolerance (i.e., 0.01\% of UB). LB is determined from the master problem, whereas UB is determined after solving subproblems.

\subsubsection{Master problem}
The formulation of the master problem is given in \equref{eq:objective function_of_master}. The master problem computes the lower bound of the planning problem by summing the investment cost and the lower approximation of the subsequent expected subproblem costs. Each $\alpha_d$ represents the subproblem cost for days classified into each day type. It is initially bounded by $\underline{\alpha}$, which is the parameter given as the lower bound for the subproblem cost. $n\in\mathcal{N}$ is the index of benders iterations. In every $n$th iteration, Benders cuts represented by $\Gamma_{d}^{(n)}$ (see \equref{eq:bender's_cut}) are added as additional constraints for days $d\in\mathcal{D}$, as shown in \equref{eq:optimality_cut}. The lowerbound of the total cost, so-called LB, is determined by the optimal solution of the master problem (i.e., $LB=\mathcal{I}_c U_l^{*}{+}\mathcal{I}_p R_{l}^{*}{+}\mathcal{I}_{e} C_{l}^{*}{+}\sum_{d\in\mathcal{D}}\alpha_d^{*}$).
\begin{align}
& \minimize_{\forall, U,C,R,\alpha} \mathcal{I}_c U_l{+}\mathcal{I}_p R_{l}{+}\mathcal{I}_{e} C_{l}{+}\sum_{d\in \mathcal{D}}\alpha_d
\label{eq:objective function_of_master}
\end{align}
%\vspace{-8pt}
{\centering \text{subjected to : \eqref{eq:available p rating}-\equref{eq:c-rate},}\par}
\vspace{-10pt}
\begin{align}
& \alpha_d \geq \underline{\alpha},~ \alpha_d \geq \Gamma_{d}^{(n)}, \ \forall d, \forall n
\label{eq:optimality_cut}
\end{align}
\subsubsection{Subproblem}
In the subproblem associated with day $d$, a daily AR-OPF model with the time-step discretization of 15 minutes evaluates the operational advantages of ESSs while considering real operational conditions. The variables with subscript $l, \phi, t$ are defined for $l \in \mathcal{L}, \phi \in \Phi_d, t \in \mathcal{T}$, respectively.
The sufficiency of the ESS allocation is assessed by checking if the uncovered dispatch error (see \equref{eq:error expression}) satisfies the LDER for day $d$ as shown in \equref{eq:error reduction rate}. As indicated in \equref{eq:active prosumption with unserved load} and \equref{eq:reactive prosumption with unserved load}, we introduce positive and negative unserved active load terms ($ ulp_{l \phi t}^{+}, ulp_{l \phi t}^{-} \in \mathbb{R}^+$) and positive and negative unserved reactive load terms ($ ulq_{l \phi t}^{+}, ulq_{l \phi t}^{-} \in \mathbb{R}^+$) to the active prosumption and reactive prosumption at bus $l$, scenario $\phi$ and time $t$, respectively. They correspond to the amount that should be curtailed from the prosumption to primarily comply with the LDER constraint along with other operational constraints, even in the case of insufficient capacity of ESSs. 
\begin{align}
& |\sum_{l\in\mathcal{L}}\epsilon_{l \phi t}|\leq\theta_{\phi t d}|\sum_{l\in\mathcal{L}}\Delta p_{l \phi t}|, \ \forall \phi, \forall t
    \label{eq:error reduction rate} \\
& p'_{l \phi t}=p_{l \phi t}{+}ulp_{l \phi t}^{+}{-}ulp_{l \phi t}^{-}, \ \forall \l, \forall \phi, \forall t
\label{eq:active prosumption with unserved load}
\\
& q'_{l \phi t}=q_{l \phi t}{+}ulq_{l \phi t}^{+}{-}ulq_{l \phi t}^{-}, \forall \l, \forall \phi, \forall t
\label{eq:reactive prosumption with unserved load} \\
& p'_{l \phi t}=\tilde{p'}_{l \phi t}{-}\Delta p'_{l \phi t}, \ \forall \l,\forall \phi, \forall t
\label{eq:prosumption prediction2} \\
& DP_{td}={\sum_{l\in \mathcal{L}}(\tilde{p_{ltd}'}{+}r_{l}\tilde{f}_{ltd})}, \ \forall t 
\label{eq:dispatch plan2} \\
& \sum_{l\in \mathcal{L}}(\Delta {p_{l \phi t}'}{+}r_{l}\Delta f_{l \phi t})=\sum_{l\in \mathcal{L}}({\epsilon}_{l \phi t}{+}{p_{l \phi t}^E}), \ \forall \phi, \forall t
\label{eq:error expression2} 
\end{align}
The AR-OPF problem embedding the dispatchability for the subproblem is formulated by replacing \equref{eq:prosumption prediction} with \equref{eq:prosumption prediction2}. In other words, we replace the prosumption $p_l$ by $p'_l$ in the relevant equations. $\tilde{p'_l}$ is employed in place of $\tilde{p_l}$ to determine the daily dispatch plan, as shown in \equref{eq:dispatch plan2}. $\Delta p_l$ of \equref{eq:error expression} is substituted with $\Delta p'_l$ to build \equref{eq:error expression2}. 
Similarly, $p'_l$ and $q'_l$ replace $p_l$ and $q_l$ in the active power balance equations formulated with the state variables (i.e., \equref{eq:complex power balance up}, \equref{eq:complex power balance down}) and the auxiliary variables (i.e., \equref{eq:lowerbound complex power balance up}, \equref{eq:lowerbound complex power balance down}, \equref{eq:upperbound complex power balance up}, \equref{eq:upperbound complex power balance down}).
The equations of AR-OPF model with ESSs are re-defined as \equref{eq:AROPF_ESS_constraints}, where $\varphi'=\{S^t,v,f,\hat{S}^t,\Bar{v},\Bar{f}, \Bar{S}^t,s'{=}(p'{+}p^{E}){+}j(q'{+}q^{E}), ulp^{+}, ulp^{-}, ulq^{+},\\ ulq^{-}\}$ is the set of variables and $\kappa=\{\mathbf{G},z,b,v^{max},v^{min},\\I^{max}, P^{max}, Q^{max}\}$ is the set of parameters. The ESS power is governed by the set of operational constraints as \equref{eq:ESS_operation}.
\begin{align}
\label{eq:AROPF_ESS_constraints}
&\Theta'(\varphi_{\phi t}',\kappa)\geq0, \forall \phi, \forall t \\
\label{eq:ESS_operation}
&\Xi(\eta_{\phi t},\xi)\geq0, \ \forall \phi, \forall t.
\end{align}  
 However, we can intuitively expect that having \equref{eq:error expression2} cannot be compliant with the mathematical formulation of the power flow equations (i.e., \equref{eq:complex power balance up}-\equref{eq:relaxed internal current}) in the case of insufficient capacity of ESS to satisfy the LDER constraint \equref{eq:error reduction rate}. 
 % (i.e., $\omega_l$ is too limited to make $\epsilon_l$ comply with the LDER constraints). 
  The insufficient power rating of ESS means that possible $p_l^E$ is small, making $\epsilon_l$ too large to comply with LDER constraint (see \equref{eq:error expression2}). However, instead of making the problem infeasible, LDER constraint and \equref{eq:error expression2} are both satisfied by reducing the total dispatch error (left-hand side of \equref{eq:error expression2}). This leads to the violation of the physical law of power flow because the grid losses error should take an unrealistic value that has the same order of magnitude as the prosumption deviation. In this way, the prosumption error and grid losses error cancel out each other to make the overall value of the left-hand side as small as the right-hand side of \equref{eq:error expression2}. It would induce the increase of the current $f_l$ such that the left-hand side of \equref{eq:relaxed internal current} becomes strictly greater than the right-hand side, which leads to the inexactness of the solution. Therefore, we introduce an iterative algorithm, \algref{algo:losses realization}, comprising two additional slack variables, $\gamma_{\phi t}^m$ and $\zeta_{\phi t}$, to replace \equref{eq:error expression2} by \equref{eq:modified error expression1} and \equref{eq:modified error expression2} such that the value of the internal current would never deviate away from the real value (i.e., the exactness of the solution is guaranteed). 
  	$\gamma_{\phi t}^m$ represents the realized grid losses deviation at $m$th iteration for scenario $\phi$ and time $t$, where $m\in\mathcal{M}$ is the index of iterations of the algorithm. $\zeta_{\phi t}$ indicates the unrealized part of the grid losses that should be updated to $\gamma_{\phi t}^m$ after each iteration for scenario $\phi$ and time $t$. $\gamma_{\phi t}^m$ achieves the accurate value of grid losses deviation, as the absolute value of $\zeta_{\phi t}$  reaches value below the defined tolerance (i.e., 1e-5 pu.).
\begin{align}
& \sum_{l\in \mathcal{L}}(\Delta p'_{l \phi t}{+}r_l\Delta f_{l \phi t})=\sum_{l\in \mathcal{L}}(\epsilon_{l \phi t}{+}p^E_{l \phi t}){+}\zeta_{\phi t}, \ \forall \phi, \forall t
\label{eq:modified error expression1}\\
& 
\setlength{\jot}{-3pt} 
\sum_{l\in \mathcal{L}}\Delta p'_{l \phi t}{+}\gamma_{\phi t}^m =\sum_{l\in \mathcal{L}}(\epsilon_{l \phi t}{+}p^E_{l \phi t}), \ \forall \phi, \forall t
\label{eq:modified error expression2}
\end{align}
\begin{algorithm}[H]
\caption{Iterative realization of grid losses deviation}
\algorithmicrequire $\theta_d (LDER), \kappa, s=p{+}jq, R^*, C^*$ (see \equref{eq:fixed})\\
\vspace{-12pt}
\begin{algorithmic}[1]
    \STATE $\textbf{Initialization : }m=1, \gamma^1=0, \zeta=1$; 
    \WHILE{$|\zeta| \geq tolerance$}
        \STATE Solve a subproblem including \equref{eq:modified error expression1} and \equref{eq:modified error expression2}
      %  \STATE $\sum_{l\in \mathcal{L}}\Delta p_{l}{+}\gamma^{m}=\sum_{l\in \mathcal{L}}(\epsilon_{l}{+}\omega_{l})$
      %  \STATE $\sum_{l\in \mathcal{L}}(\Delta p_{l}{+}r_l\Delta f_{l})=\sum_{l\in \mathcal{L}}(\epsilon_{l}{+}\omega_{l}){+}\zeta$
        \STATE $\gamma^{m{+}1} \gets \gamma^{m}{+}\zeta$  
        \STATE $m \gets m{+}1$
    \ENDWHILE
%    \RETURN $\gamma^m, p^{E}$
\end{algorithmic}
\label{algo:losses realization}
\end{algorithm}
Finally, the subproblem is described with an objective of minimization of the total grid losses and unserved load to satisfy the LDER constraint, and the operation period spans all days grouped into each day-type over the planning horizon.

\begin{align}
\begin{aligned}
\begin{split}
\minimize_{\forall \varphi',\eta,UL} :  \,\mathcal{SC}_d = Y N_d \sum_{t \in \mathcal{T}}\sum_{\phi \in \Phi_d}\lambda_{\phi d}(w_l\sum_{l\in \mathcal{L}}r_{l}f_{l \phi t}\\
+w_u\sum_{l\in \mathcal{L}}(ulp_{l \phi t}^{+}+ulp_{l \phi t}^{-}+ulq_{l \phi t}^{+}+ulq_{l \phi t}^{-}))
\end{split}
\label{eq:objective function_2nd block}
\end{aligned}
\end{align}
\vspace{-10pt}
{\centering subject to: \equref{eq:error reduction rate}, \equref{eq:active prosumption with unserved load}, \equref{eq:reactive prosumption with unserved load} \equref{eq:dispatch plan2}, \equref{eq:AROPF_ESS_constraints}-\equref{eq:modified error expression2},\par}
\begin{align}
& R_l = {R}_l^{*} : \mu_{ld}, ~ C_l = C_l^{*} : \vartheta_{ld},\ \forall l.
\label{eq:fixed}\\
& \Gamma_{d}^{(n)}=\big[\mathcal{SC}_d^{*}{-}\sum_{l\in \mathcal{L}}(\mu_{l d}(R_l{-}R_l^{*}){-}\vartheta_{l d}(C_l{-}C_l^{*}))\big], \forall d, \forall n
\label{eq:bender's_cut}
\end{align}
where $UL=\{ulp^{+},ulp^{-}, ulq^{+},ulq^{-}\}$ is the set of variables related to the unserved load. $w_l$ and $w_u$ are the weight coefficients associated with the grid losses minimization and unserved load, respectively. \eqref{eq:fixed} describes that the ESS power ratings and the energy reservoirs are fixed to the optimal solution values of the master problem. $\mu_{l,d}$ and $\vartheta_{l,d}$ are the duals of constraints related to the fixed ESS capacities, and are used to form the benders cuts for the master problems as shown in \equref{eq:bender's_cut}. UB is calculated summing the optimal investment cost and the subproblem costs (i.e., $UB=IC^{*}{+}\sum_{d\in\mathcal{D}}\mathcal{SC}_d^{*}$).
\section{Simulation}
\label{sec:simulation}
\vspace{-2pt}
We validate the performance of the proposed methods with an existing Swiss distribution network with 55 bus and large capacity of renewable generation, as shown in \figref{fig:Aigle grid}. The base voltage is 21kV and the base 3 phase power is 6MVA. 2.7MWp of PV generation capacity and 805kVA of hydropower generation capacity is installed. The planning horizon is set as 10 years, and we assume that the load consumption does not grow over the planning horizon. According to the indications of the operator of this grid, the number of candidate nodes for ESS installation is set as 5 out of 55 nodes (see \tabref{tb:Simulation Parameters}). To make sure to achieve the dispatchability, the penalty cost for the dispatch error is assumed as \$700/MWh, which is significantly higher than a typical price settled in energy markets. The optimization problems are solved using the solver MOSEK via the MATLAB interface YALMIP. %on a PC with Intel Core i7-7700HQ CPU 2.8GHz and 32GB RAM. %This assumption is practically reasonable because only a limited number of nodes are available to install ESS due to various geographical constraints. 
\begin{table}[ht]
	\centering
	\caption{ESS parameter and candidate nodes for simulation}
	\begin{tabular}{c|c|c|c}
		\hlinewd{1pt}
		\makecell{Maximum power\\ rating per site} & 3MW & \makecell{Maximum energy\\ reservoir per site} & 4MWh       \\ \hline
		\makecell{Installation cost\\ for energy reservoir}                          & \$300/kWh                                           & \makecell{Installation cost\\ for power rating}     & \$200/kVA \\ \hline
		\multicolumn{2}{c|}{Capital investment cost per site} & \multicolumn{2}{c}{\$0.1M} \\ \hline		
		\multicolumn{2}{c|}{Candidate nodes for ESS} & \multicolumn{2}{c}{4, 16, 27, 41, 45} \\ \hlinewd{1pt}
	\end{tabular}
	\label{tb:Simulation Parameters}

\end{table}
	\vspace{-5pt}
\begin{figure}[ht]
	\centering
	\includegraphics[scale=0.45]{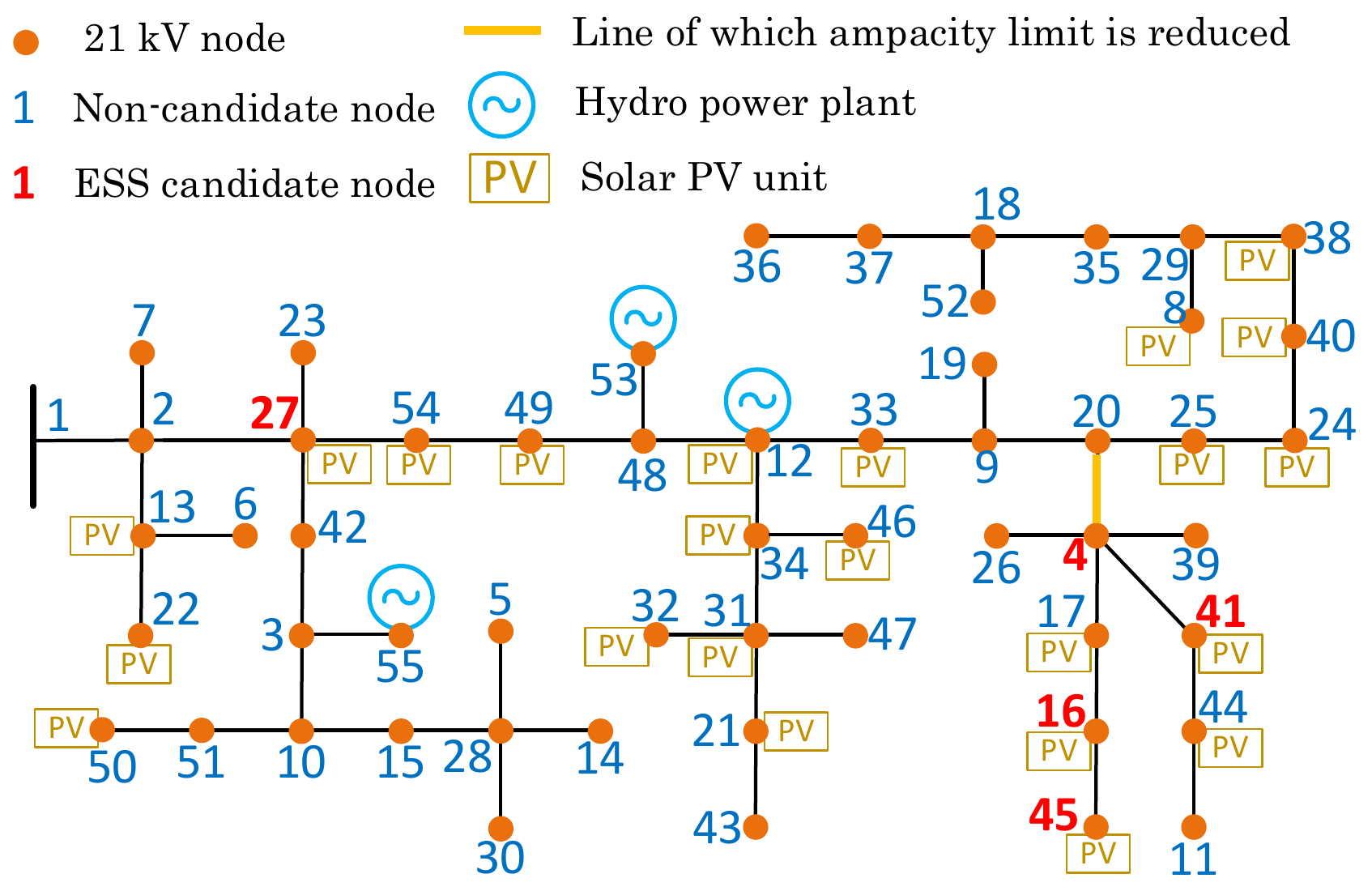}
	 \vspace{-5pt}
	\caption{Considered real 55 bus distribution feeder.}
	     
	\label{fig:Aigle grid}
\end{figure}

We have considered 8 typical day-types to cover the seasonal variation of the prosumption over the year. For each day-type, we assumed that prosumption forecasts for the simulation have been given from a reliable forecasting methodology. 1000 prosumption scenarios for each day are generated with equal probabilities based on the assumption that the prosumption profile follows a normal distribution as discussed in \secref{sec:achieving dispatchability}. Then, we applied a K-medoids clustering \cite{jain1988algorithms} based scenario reduction technique to mitigate the computational burden. The number of reduced number of scenarios is determined by an algorithm explained in \secref{appendix:scenario}. 
\subsection{Planning with 1 day under hourly dispatch}
To illustrate the role of the 1st and 2nd block of the problem, we demonstrate the result of the simplified simulation considering an hourly dispatch for 1 representative day in two cases: case 1 with the original ampacity of the lines specified for the given grid and case 2 with the ampacity of line between node 4 and 20 reduced from the original value (see \figref{fig:Aigle grid}). 32 representative scenarios obtained from the scenario reduction algorithm in \secref{appendix:scenario} are used as operating scenarios. \tabref{tb:installation_result1} indicates the optimal allocation result of ESS. The optimal result from the 1st block specifies the capacity in size of power rating and energy reservoir of ESS. However, the determined allocation of ESS cannot be guaranteed to be feasible and optimal to satisfy the LDER for the real operation of the grid,  since the grid losses and the ampacity constraint were neglected in the 1st block problem. After the 2nd block of the problem, as shown in the result of case 1 in \tabref{tb:installation_result1}, the optimal site of the ESS considering the objective of minimization of the grid losses is determined as node 4 and 27, resulting in the reduction of the grid losses and unserved energy compared to the result of the 1st block problem. 

In case 2, we can observe that a part of the load consumption was unserved to satisfy the LDER in the condition of restrained ampacity limit with the determined ESS allocation from the 1st block. In this regard, the result of the 2nd block shows the change in the allocation of the ESS due to the bottleneck of the line. The power rating and energy reservoir of the ESS on Node 4 is reduced, and those of the ESS on Node 27 is increased. \tabref{tb:simplified_result} shows that the unserved energy decreased to near zero after re-allocating the ESS.

\begin{table}[h]
	\centering
\caption{ESS allocation result}
\label{tb:installation_result1}
	\begin{tabular}{c|c|ccc}
		\hlinewd{1pt}
		\multirow{2}{*}{Case} & Problem                    & Location & Power rating & Energy reservoir \\ \cline{2-5}
		& 1st Block                  & 41       & 712 kVA      & 1.958 MWh        \\ \hline
		\multirow{2}{*}{1}    & \multirow{2}{*}{2nd Block} & 4        & 372 kVA      & 803 kWh          \\
		&                            & 27       & 338 kVA      & 1.176 MWh        \\ \hline
		\multirow{2}{*}{2}    & \multirow{2}{*}{2nd Block} & 4        & 360 kVA      & 732 kWh          \\
		&                            & 27       & 350 kVA      & 1.247 MWh                 \\ \hlinewd{1pt}    
	\end{tabular}
\end{table}

	\vspace{-5pt}
\begin{table}[h]
\centering
    \caption{Comparison between the result of 1st block and 2nd block}
\begin{tabular}{l|l|l|l|l}
\hlinewd{1pt}
Case                      & Horizon                   & Type of cost                                                                                 & \begin{tabular}[c]{@{}c@{}}Allocation \\ of 1st block\end{tabular} & \begin{tabular}[c]{@{}c@{}}Allocation \\ of 2nd block\end{tabular} \\ \hlinewd{1pt} 
\multirow{4}{*}{case1}                                            & \multirow{2}{*}{10 years} & Investment cost                                                                  & \$0.830M                                                                      & \$0.935M                                                                      \\ \cline{3-5} 
                                            &                           & Penalty cost                                                                     & \$0.708M                                                                     & \$0.708M                                                                      \\ \cline{2-5} 
                                            & \multirow{2}{*}{1 year}   & \begin{tabular}[c]{@{}l@{}}Unserved energy \\ to satisfy the LDER\end{tabular} & 13.53kWh                                                                      & 8.68kWh                                                                      \\ \cline{3-5} 
                                            &                           & Grid losses                                                                      & 41.736MWh                                                                      & 39.708MWh                                                                      \\ \hlinewd{1pt}
\multicolumn{1}{c|}{\multirow{4}{*}{case2}}                       & \multirow{2}{*}{10 years} & Investment cost                                                                  & \$0.830M                                                                      & \$0.936M                                                                      \\ \cline{3-5} 
\multicolumn{1}{c|}{}                       &                           & Penalty cost                                                                     & \$0.708M                                                                      & \$0.708M                                                                      \\ \cline{2-5} 
\multicolumn{1}{c|}{}                       & \multirow{2}{*}{1 year}   & \begin{tabular}[c]{@{}l@{}}Unserved energy\\ to satisfy the LDER\end{tabular}     & 17.407MWh                                                                      & 8.11kWh                                                                      \\ \cline{3-5} 
\multicolumn{1}{c|}{}                       &                           & Grid losses                                                                      & 41.623MWh                                                                      & 39.714MWh                                                                      \\ \hlinewd{1pt}
\end{tabular}
    \label{tb:simplified_result}
\end{table}
     
	\vspace{-5pt}
\subsection{Planning with full scenarios under 15 min dispatch}
The proposed planning procedure is applied to the full set of scenarios with 8 typical days under 15 min interval dispatch. We apply scenario reduction technique to 1000 scenarios of each day-type based on the algorithm in \secref{appendix:scenario}. The largest number of reduced scenarios over all day-types is determined as 39 scenarios. \tabref{tb:installation_result2} shows the optimal ESS locations and sizes. We exhibit the results for two cases corresponding to case with no ESS installed and case with the optimal allocation of ESS. \figref{fig:Dispatch_plan} illustrates the operation result for day-type 1, showing the prosumption prediction considering 39 scenarios of the prosumption profiles, the dispatch plan, and the active power infeed through GCP corresponding to each scenario in the case of no ESS (see \figref{fig:Dispatch_plan}.(a)) and the optimal ESS allocation (see \figref{fig:Dispatch_plan}.(b)). The dispatch result without ESS shows that the dispatch error is significant, especially in the time where the production from PV is high.  
On the other hand, in the case with the optimal ESS allocation, the active power infeed of every prosumption scenario follows the dispatch plan with small error. The cost analysis between the case with ESS and without ESS in \tabref{tb:cost_result} demonstrates quantitatively the capability of ESS to handle uncertainties within the grid. The annual dispatch error of the case without ESS is about 22 times of that in the case with ESS. The difference in the dispatch error is translated into the significant gap in the total cost for 10 years of operation: \$9.09 Million with the default system configuration, and \$1.41 Million with the optimal ESS allocation. Consequently, this result demonstrates the advantages for the DSO to invest on ESS in view of their technical and economical profit.
%The dispatch plan follows the total predicted prosumption considering the grid losses, which is the average point over the slack active powers of all prosumption profiles. 

%The difference in the penalty costs of two cases represents that the annual dispatch error of the distribution grid without ESS is about 9 times larger than that of the grid with ESS.

\begin{table}[h]
\centering
\caption{ESS allocation result}
\begin{tabular}{c|ccc}
\hline
Problem   & Location & Power rating & Energy reservoir \\ \hline
1st Block & 16       & 1.312MVA    & 1.816MWh         \\ \hline
2nd Block & \makecell{4\\27}        & \makecell{544.55kVA\\764.42kVA}    & \makecell{908.82kWh\\921.65kWh}         \\ \hline
\end{tabular}
\label{tb:installation_result2}
\end{table}
	\vspace{-5pt}

\begin{table}[h]
\centering
\caption{Cost and Operational advantage comparison}
\begin{tabular}{l|lcccc}
\hline
Horizon                   &                                                                                                          & Without ESS &  With ESS \\ \hline
\multirow{3}{*}{10 years} & Total cost                                                                                               & \$9.09M                         & \$1.41M                      \\ 
                          & Investment cost                                                                                          & -                               & \$1.01M                      \\
                          & Penalty cost                                                                                             & \$9.09M                         & \$0.40M                      \\ \hline
\multirow{3}{*}{1 year}   & Uncovered error & 1.299GWh                       & 57.38MWh    
\\
                          & Grid losses                                                                                    & 59.578MWh                        & 53.879MWh                     \\
                          & Total energy consumed                                                                                    & 8.101GWh                        & 8.139GWh                     \\ \hline
\end{tabular}
\label{tb:cost_result}
\end{table}
	\vspace{-5pt}
    \begin{figure}[ht]
    	\centering
    	\renewcommand{\thesubfigure}{}
    	\subfigure[(a)]
    	{\includegraphics[width=0.88\linewidth]{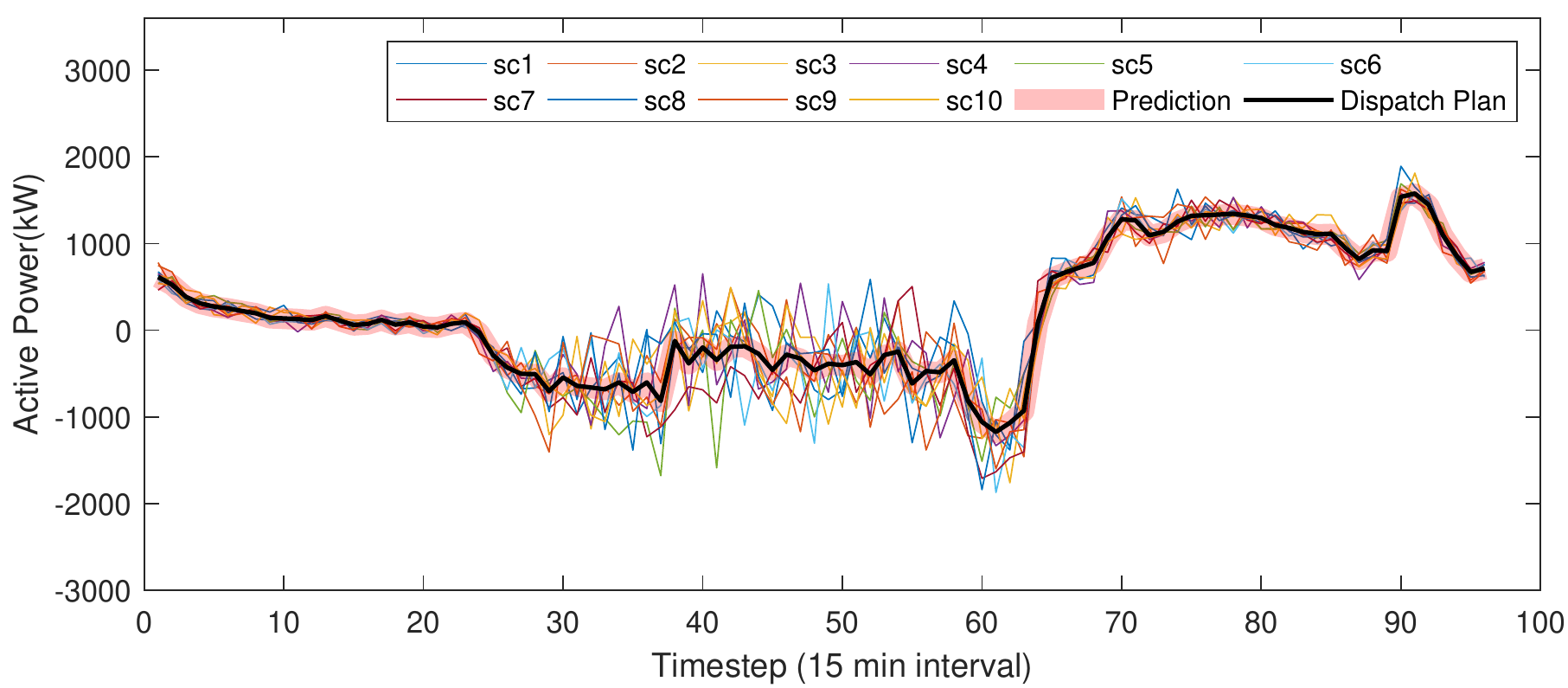}}\\
    	 \vspace{-10pt}
    	\subfigure[(b)]
    	{\includegraphics[width=0.88\linewidth]{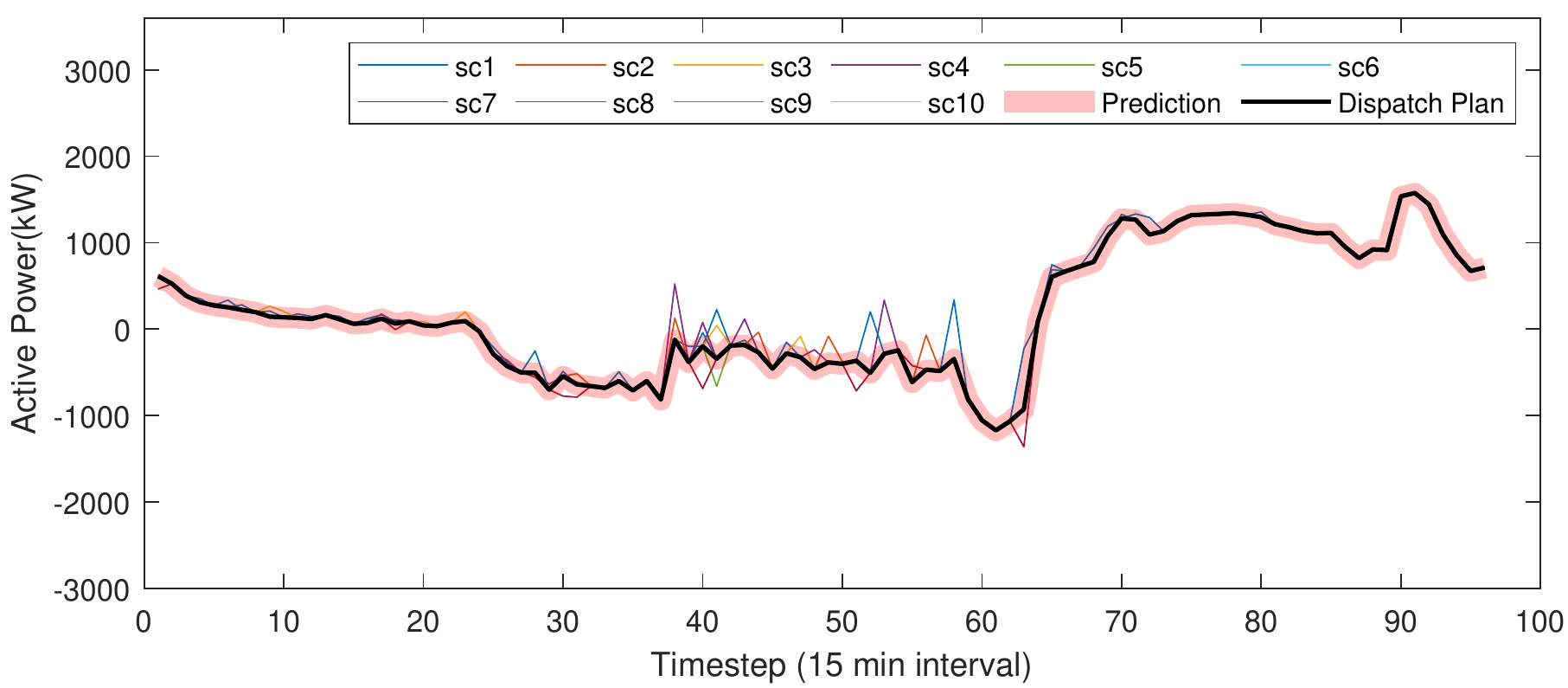}}
     	 \vspace{-10pt}	
    	\caption{Prosumption prediction, dispatch plan and active power through GCP in each scenario: (a) Day 1(No ESS), (b) Day 1(With ESS).}
    	\label{fig:Dispatch_plan}
    \end{figure}
	\vspace{-5pt}
%\subsection{Quantification of the compression of the solution space associated with the conservative constraints of AR-OPF}

%\begin{comment}
\section{Discussion on the limitation of the research}
Modeling the prosumption uncertainty can affect the fidelity of the optimal solution regarding the allocation of ESS and the dispatchabiliy level of an ADN. However, as modeling of prosumption stochasticity is not the scope of this paper, the prosumption scenarios are generated simply assuming that the prosumption follows a normal distribution. Nevertheless, when a modeler is equiped with a robust scenario generater which models accurately the prosumption uncertainty, the proposed approach can guarantee the reliable performance. Another limitation of the proposed approach relates to the condition on the objective function for guaranteeing the exactness of the solution. This condition may limit the extendibility of the application for various control objectives of DSO's interest. For example, the minimization of the voltage deviation or control on ESS energy level may not satisfy the condition for the exactness, and having these objective terms within the objective function may affect the quality of the solution. Therefore, appropriate modification on the AR-OPF model would be necessary corresponding to such control objectives.
%\end{comment}

\section{Conclusion}
In this study, we have presented a tool for the optimal planning of ESSs within a distribution network to achieve its dispatchability. We have shown that the uncertainty of the prosumption can be compensated sufficiently with the allocation and exploitation of ESS. The non-approximated and convex OPF model, or the AR-OPF model is implemented to account for the operational conditions of the distribution network accurately. The planning problem is decomposed into two blocks to satisfy the condition for the exactness of the solution via the AR-OPF model. In the 1st block, the allocation of ESS is determined along with the corresponding LDER by implementing the linearly approximated OPF model. The AR-OPF is used in the 2nd block of the problem to check the compatibility of the allocated capacity for the real operation of the grid to satisfy the LDER and to determine the optimal location of the ESS to minimize the grid losses. 
We validated the effectiveness of the proposed method for a real Swiss ADN of 55 nodes by demonstrating that the allocation of ESS successfully reduced the dispatch error.
\begin{appendices}\section{Scenario reduction algorithm}
\label{appendix:scenario}
%Scenario reduction techniques are typically applied to select a subset of initial scenario set. 
%The quality of a stochastic optimization solution highly depends on how much the scenarios in the subset can properly preserve the stochasticity of the origianl scenario set. In this respect, we propose an algorithm as shown in \algref{algo:scenario reduction} to determine the minimum number of scenarios to represent the total scenarios. 
Each of the generated scenarios is expressed as vector $\bm{\nu}_\phi=\{\mathbf{\nu}^{1}_\phi,...,\mathbf{\nu}^{H}_\phi\}, \forall \phi \in \Phi_d$ where $H=\sum_{l \in \mathcal{L}} (T^{AL(l)}{+}T^{RL(l)}\\{+}T^{PV}$. $T^{AL(l)}, T^{RL(l)},$ and $T^{PV}$ are the time duration of the daily active, reactive load at node $l$, and PV irradiation profiles, respectively. Firstly, we construct a cumulative distribution function (CDF) for each random variable $\bm{\nu}^{h}$ with scenario set $\Phi_d$, which is given by $y_{\phi}^{h}=cdf(\nu^{h}_{\phi})=P(\bm{\nu}^{h}\leq \nu^{h}_{\phi}), \forall \phi \in \Phi_d, h \in \{1,...,H\}=\mathcal{H}$. $cdf^{-1}(y_{\phi}^{h}), y_{\phi}^{h} \in [0,1]$ represents its inverse function.
%that corresponds to random variable $\bm{\nu}^{h}$ with scenario set $\Phi_d$. 
The obejctive of the algorithm is to find the minimal number of reduced scenarios such that the average distance between the CDF of the initial scenario set and that of the reduced scenario set over number of check points becomes smaller than a given tolerance. We define the check points for caclulating the distance between the CDF curve of scenario set $\Phi_d$ with another CDF curve as $cdf^{-1}(y_{\Phi_d}(q)), y_{\Phi_d}(q) \in [0,1], \forall q \in \{1,...,N_q\}$. In this paper, we selected $N_q=5$, with quantiles ranging $[0.05, 0.95]$.  
 The scenario reduction process is initialized by applying K-Medoids clustering method \cite{jain1988algorithms}  based on the Euclidean distance between each scenario pairs to reduce the original scenario set into set $\Phi_d'$ with a single representative scenario. The average distance between CDFs of different scenario set $\Phi_d$ and $\Phi_d'$ is calculated by index defined by $\Delta=\sum_{q}\omega_q\sqrt{\frac{1}{H}\sum_{h} \left(\frac{cdf^{-1}(y_{\Phi_d}^{h}(q)){-}cdf^{-1}(y_{\Phi_d'}^{h}(q))}{cdf^{-1}(y_{\Phi_d}^{h}(q))}\right)^2}$, where $\omega_q$ is the weight coefficient assigned to $q$th check point. As the distance is bigger than a threshold value, %(e.x., 0.08, which corresponds to 8 percent of the sample values of the original scenario set),
  the scenario reduction is re-applied to produce a scenario set with incremented number of scenarios than the previous iteration.
  \figref{fig:scenarios} shows a CDF of original scenario set with the CDFs of different reduced scenario sets regarding a single random variable. The proposed algorithm determines that the minimum number of scenarios required is 39. \figref{fig:diff_scenarios} shows the evlolution of $\Delta$ w.r.t the number of reduced scenarios.
  %$\Delta=\frac{1}{H}\sum_{h}\sqrt{\sum_{q} \omega_q\left(\frac{(cdf^{-1}(y_{\Phi_d}^{h}(q)){-}cdf^{-1}(y_{\Phi_d'}^{h}(q)))}{cdf^{-1}(y_{\Phi_d}^{h}(q))}\right)^2}$
  \begin{figure}[H]
	\hspace{-1em} 
	\subfigure[]{% 
		\includegraphics[width=.25\textwidth]{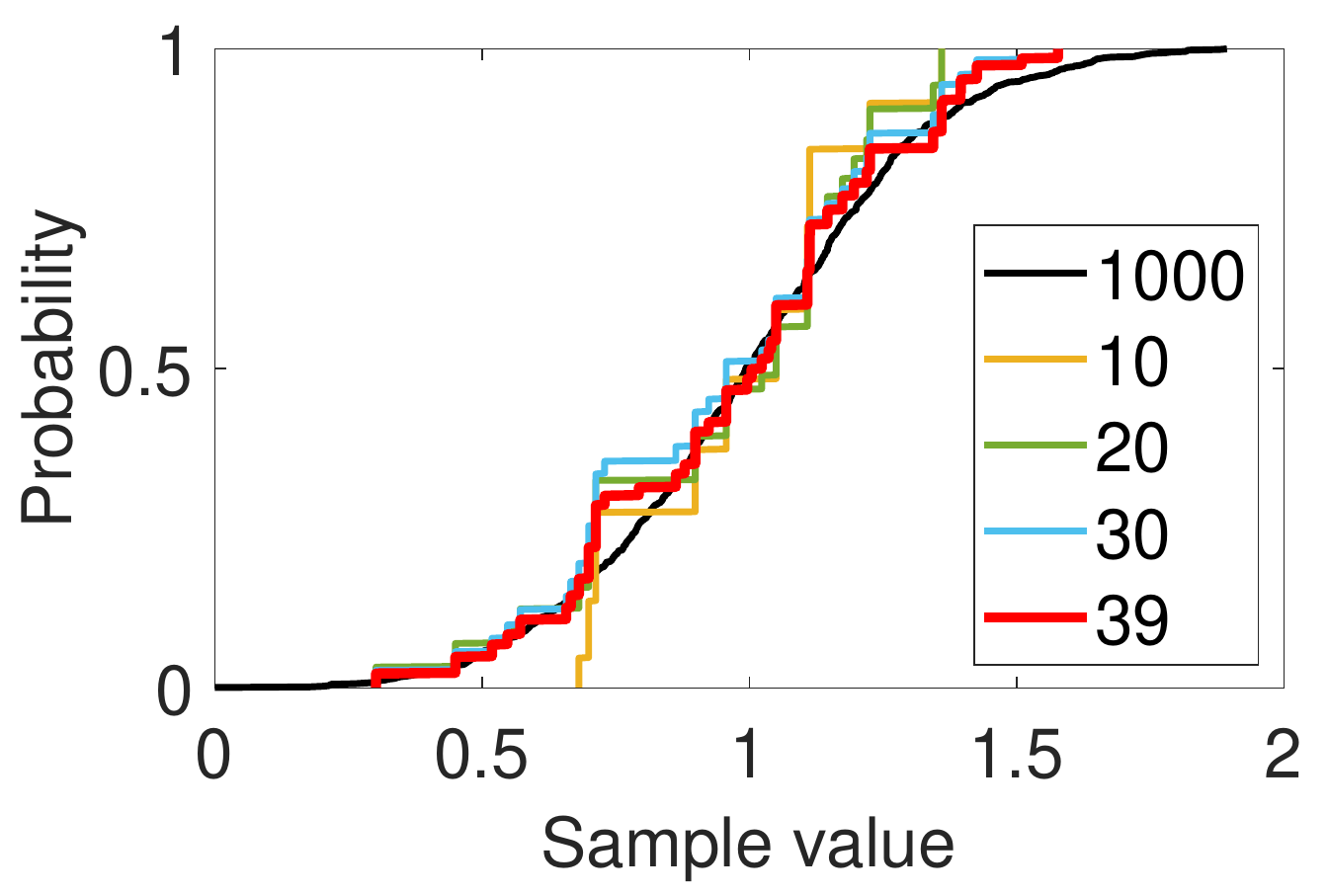} \label{fig:scenarios} 
	}
\hspace{-1em}
	\subfigure[]{% 
		\includegraphics[width=.25\textwidth]{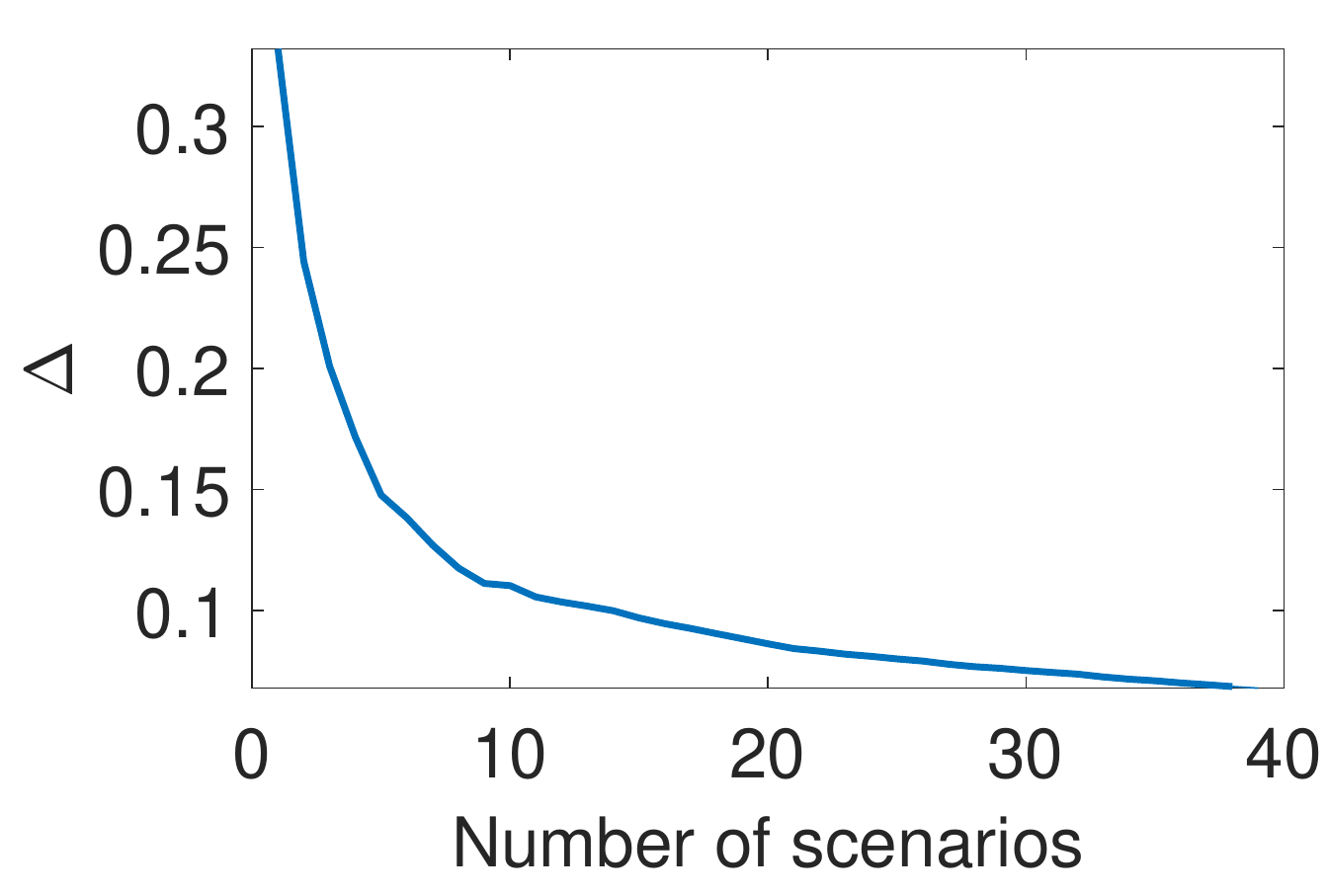} \label{fig:diff_scenarios} 
	} 
\hspace{-1em}
\vspace{-1em}
	\caption{(a) CDF of initial scenario set and reduced scenario sets with different number of reduced scenarios, (b) Average of normalized distance between initial and reduced scenario sets with different number of reduced scenarios} 
	%Comparison between CDFs of initial scenario set and reduced scenario set: 
\end{figure}
\vspace{-0.5em}
\begin{algorithm}[H]
	\caption{Scenario reduction}
	%	\algorithmicrequire $\theta_d (LDER),  s=p{+}jq$ , $R^*$ (see \equref{eq:fixed_power}), $C^*$ (see \equref{eq:fixed_energy})\\
	%	\vspace{-12pt}
	\begin{algorithmic}[1]
		\STATE Generate $N_{\Phi_d}$ scenarios for $\bm{\nu}$
		\STATE Draw CDF graphs of $\bm{\nu}^{h}, \forall h \in \mathcal{H}$ with scenario set $\Phi_d$. %\forall \phi \in \Phi$ 
		\STATE \textbf{Initialization: }$N_{\Phi_d'}=0, \Delta=\infty$
		\WHILE{$|\Delta| \geq tolerance$}
		\STATE $N_{\Phi_d'} \gets N_{\Phi_d'}+1$  
		\STATE Obtain $N_{\Phi_d'}$ scenarios for $\bm{\nu}$ via K-medoids clustering 		
		\STATE Draw CDF graphs of $\bm{\nu}^{h}, \forall h \in \mathcal{H}$ with scenario set $\Phi_d'$. %\forall \phi \in \Phi$ 
		\STATE Compute $\Delta$ between the two CDFs of set $\Phi_d$ and $\Phi_d'$.
		\ENDWHILE
		\RETURN $N_{\Phi_d'}, \bm{\nu}_{\phi_{d}'}, \forall \phi_d'\in\Phi_d' $
	\end{algorithmic}
	\label{algo:scenario reduction}
\end{algorithm}

\end{appendices}

\newcommand{\BIBdecl}{\setlength{\itemsep}{0.25 em}}
\bibliographystyle{IEEEtran}
\bibliography{bibtex/bib/IEEEexample}

\end{document}